\documentclass[doublespace,times,10pt]{article}

\usepackage{authblk}
\usepackage{amsmath}
\usepackage{rotating}
\usepackage[hidelinks,breaklinks=true]{hyperref}
\usepackage{color}
\usepackage{enumitem}

\title{Using phase II data for the analysis of phase III studies: an application in rare diseases}
\author[1,*]{Simon Wandel}
\author[1]{Beat Neuenschwander}
\author[2]{Christian R\"over}
\author[2]{Tim Friede}
\affil[1]{Novartis Pharma AG, Basel, Switzerland}
\affil[2]{Department of Medical Statistics, University Medical Center G\"ottingen, Humboldtallee 32, 37073 G\"ottingen, Germany}
\date{}

\begin{document}
\raggedright
\maketitle

\textbf{Short title:} Using phase II data in phase III\\[2\baselineskip]
\textbf{Word count (excluding formulas, tables, figures):} 3578\\[2\baselineskip]
\textbf{Funding:} This research has received funding from the EU's 7th Framework Programme for research, technological
development and demonstration under grant agreement number FP HEALTH 2013-602144 with project title
(acronym) "Innovative Methodology for Small Populations Research" (InSPiRe).\\[2\baselineskip]
\textbf{*Corresponding author:} Simon Wandel, Novartis Pharma AG, Basel
\clearpage

\section*{Abstract}
\subsection*{Background}
% 62 words
Clinical research and drug development in orphan diseases is challenging, since large-scale randomized studies are difficult to conduct. Formally synthesizing the evidence is therefore of great value, yet this is rarely done in the drug approval process. Phase III designs that make better use of phase II data can facilitate drug development in orphan diseases.

\subsection*{Methods}
% 122 words
A Bayesian meta-analytic approach is used to inform the phase III study with phase II data. It is particularly attractive, since uncertainty of between-trial heterogeneity can be dealt with probabilistically, which is critical if the number of studies is small. Furthermore, it allows quantifying and discounting the phase II data through the predictive distribution relevant for phase III. A phase III design is proposed which uses the phase II data and considers approval based on a phase III interim analysis. The design is illustrated with a non-inferiority case study from an FDA approval in herpetic keratitis (an orphan disease). Design operating characteristics are compared to those of a traditional design, which ignores the phase II data.

\subsection*{Results}
% 97 words
An analysis of the phase II data reveals good but insufficient evidence for non-inferiority, highlighting the need for a phase III study. For the phase III study supported by phase II data, the interim analysis is based on half of the patients. For this design, the meta-analytic interim results are conclusive and would justify approval. In contrast, based on the phase III data only, interim results are inconclusive and would require further evidence.

\subsection*{Conclusions}
To accelerate drug development for orphan diseases, innovative study designs and appropriate methodology are needed. Taking advantage of randomized phase II data when analyzing phase III studies looks promising because the evidence from phase II supports informed decision making. The implementation of the Bayesian  design is straightforward with public software such as \texttt{R}.\\[2\baselineskip]
\textbf{Keywords:} drug development in rare diseases; phase III studies; Bayesian statistics; meta-analysis 
\clearpage

\section*{Introduction}
\label{sec:introduction}
Clinical research in orphan diseases is challenging. It is often impossible or unethical to conduct large scale randomized controlled trials, which implies that only limited evidence is available for decision making. Also, shortcomings in the methodological approaches to evaluate medical products in rare diseases have been identified (e.g. Unkel et al~\cite{Unkel2016}). Whilst these problems have been recognized for some time (see Orphan Drug Act from 1983~\cite{oda1983}), only in the past few years strong efforts have been made to address them. Examples include the draft guidance by the Food and Drug Administration (FDA) for drug development in rare diseases~\cite{FDAGuidanceRD} and the latest funding scheme for rare diseases by the European Union's Horizon 2020 research program~\cite{HorizonRD}. These activities have led to intensified rare diseases research and drug development by pharmaceutical companies~\cite{Sharma2010}.\\[2\baselineskip]

With regards to the drug approval process, some flexibility on study designs and endpoints has been observed for drugs with an orphan indication~\cite{Sasinowski2012, Frank2014}. Surprisingly, however, a formal combination of the evidence (for example a meta-analysis) is rarely presented in approval dossiers. Typically, efficacy is assessed based on confirmatory trials only, meaning that other evidence (such as phase II studies) is viewed as supportive only. This poses a problem to both, regulators in charge of approving drugs and companies developing them, since it limits the evidence base for a quantitative assessment of the treatment effect. Furthermore, the combination of data reveals its power particularly in situations with limited data at hand, which is often the case for rare diseases.\\[2\baselineskip]

These challenges call for approaches to study design and analysis that allow a more efficient use of the available data, as stipulated e.g. in the 21\textsuperscript{st} Century Cures Act~\cite{21CuresAct}. The nature of the problem lends itself to the Bayesian approach. The usefulness of the Bayesian approach when meta-analyzing few (small) studies has been discussed elsewhere (see Friede et al~\cite{Friede2016a, Friede2016b}). Here, we extend the idea to incorporate existing evidence for the parameter of interest, the treatment effect corresponding to the phase III study, via a meta-analysis. This is based on concepts discussed by Spiegelhalter et al~\cite{Spiegelhalter2004}, Neuenschwander et al~\cite{Neuenschwander2010, Neuenschwander2016}, Schmidli et al~\cite{Schmidli2014} and some ideas in Gerss and K\"opcke \cite{delaPaz2010}.\\[2\baselineskip]

The paper is organized as follows. We first describe the statistical methodology, then illustrate the design using data from an FDA approved drug, and conclude with a discussion.\\[2\baselineskip]

\section*{Methods}
\label{sec:methods}
\newcommand\codata{\emph{co-data} }
\newcommand\thstar{\theta_{\star}}
\newcommand\thJ{\theta_{\cal{J}}}
\newcommand\Jstar{ {\cal{J}_{\star}}}
\newcommand\J{{\cal{J}}}
\newcommand\Ystar{Y_{\star}}
\newcommand\YJ{Y_{\cal{J}}}
\newcommand\MAC{\emph{MAC} }
\newcommand\MAP{\emph{MAP} }

\subsection*{Hierarchical models}
Hierarchical models (HM) are widely used when data are available from
more than one trial.  The models have two components: a data model and
a parameter model. Data $Y_j$ from trial $j=1,\ldots,J$ follow a
distribution $F$ parameterized by trial-specific parameters
$\theta_j$
\begin{equation}
\label{data:model}
Y_j \vert \theta_j \sim F(\theta_j)
\end{equation}
and trial parameters $\theta_j$ follow a distribution $G$
\begin{equation}
\label{parameter:model}
\theta_j \vert \eta \sim G(\eta)
\end{equation}
Inference for trial parameters can be done in a classical or Bayesian
way. The simplest hierarchical model assumes (approximately) normal
data. Often, the $Y_j$ are parameter estimates rather
than individual data. For this case, the normal-normal hierarchical
model (NNHM) is widely used:
\begin{equation}
\label{EX}
Y_j \vert \theta_j \sim N(\theta_j,s_j^2)
\end{equation}
and
\begin{equation}
\label{EX:normal}
\theta_j \vert \mu,\tau \sim N(\mu,\tau^2)
\end{equation}
For fixed standard errors $s_j$ and known (assumed) $\tau$, classical
and Bayesian conclusions for $\mu$ and trial parameters $\theta_j$ are
the same if a non-informative (improper) prior for $\mu$ is used. For
precision (inverse-variance) weights $w_j$, total precision $w_+$, and
shrinkage parameters $B_j$
\begin{equation}
w_j = \frac{1}{s_j^2+\tau^2}, \qquad w_+ = \sum_{j=1}^Jw_j,  \qquad B_j = \frac{s_j^2}{s_j^2+\tau^2}
\end{equation}
the posterior distribution of $\mu$ based on $Y=(Y_1,\ldots,Y_J)$ is
\begin{equation}
\label{NNHM:mu}
\mu \vert Y, \tau \sim N(\sum w_jY_j/w_+,1/w_+)
\end{equation}
The posterior distributions of the trial parameters $\theta_j$ are
\begin{equation}
\label{NNHM:theta}
\theta_j \vert Y,\tau \sim N(B_j\hat{\mu}+(1-B_j)Y_j,
B_j(\tau^2+B_j/w_+))
\end{equation}
where $\hat{\mu}$ is the posterior mean in (\ref{NNHM:mu}). Classical
analogues to the posterior means and standard deviations are
maximum-likelihood estimates and their standard errors. The special
cases of complete pooling and stratification arise for $\tau=0$ and
$\tau=\infty$, respectively.\\[2\baselineskip]

Intermediate values of $\tau$ lead to
different degrees of information sharing across trials, with  
the desirable properties one expects from
an approach aiming to improve inference by 
borrowing information from similar trials:
\begin{itemize}
\item The hierarchical model shrinks the trial estimates towards the
  estimate of $\mu$, which acts as a
  safeguard against over-interpreting extreme (good or bad) trial results.
  Shrinkage depends on trial size and between-trial heterogeneity. For
  large trials (small
  $s_j$), shrinkage is small, and notable
  shrinkage is only possible if $\tau$ is of small to moderate size.
\item
The hierarchical model improves precision. Since
\begin{equation}
\label{HM:var}
B_j(\tau^2+\frac{B_j}{w_+}) = 
s_j^2 - s_j^2B_j(1-\frac{w_j}{w_+})
\end{equation}
the variance in (\ref{NNHM:theta}) is always smaller than the variance $s_j^2$ of $Y_j$.
\end{itemize}

\subsection*{Between-trial heterogeneity}
The degree of between-trial heterogeneity (standard
deviation $\tau$ in (\ref{EX:normal})) for
the parameters $\theta_1,\ldots,\theta_J$ depends on the parameter
scale and the outcome standard deviation $\sigma$ for one observation
unit (for example one subject or one event). Table
\ref{tab:heterogeneity} shows four typical heterogeneities and
respective $\tau$ values for $\sigma=2$, which is 
often used as the reference standard deviation for
normal approximations of binomial, count, and survival data~\cite{Spiegelhalter2004}. 
For the four heterogeneities, Table
\ref{tab:heterogeneity} shows the range of parameter values expressed
as the ratio between the 97.5\% quantile and the median; for example,
$\tau=1$ implies a ratio of 7.1, which is clearly large and will be
rare in practice. \\[2\baselineskip]

For the common case of few trials, the size of between-trial
heterogeneity is usually highly uncertain because $\tau$ cannot be
inferred well from the data.  Therefore, it is important to use prior
distributions covering plausible $\tau$ values. Half-normal,
half-Cauchy, and half-t distributions have been suggested in this
context~\cite{Spiegelhalter2004, Gelman2006, Polson2012}.  For the
log-risk ratios used in the application, we will consider
half-normal distributions~\cite{Spiegelhalter2004} with scale
parameters 0.5 and 1, which have medians (95\%-intervals) equal to
0.34 (0.016,1.12) and 0.67 (0.031,2.24), respectively. Since $\tau = 1$ represents large heterogeneity, both priors
are weakly informative, covering small to large
heterogeneity and leaving small probabilities to unrealistically large
heterogeneities, whereby the latter prior (with median 0.67) is rather
conservative. For these priors, the 97.5\% quantile to median ratio for 
risk ratio (RR) trial parameters is 2.98 and 8.89, respectively (see Appendix).

\subsection*{Meta-analytic-predictive (MAP) prior }
When designing a new trial with parameter   $\theta_{\star}$, the predictive distribution based 
on previous data   $Y_1,\ldots,Y_J$ constitutes the prior distribution for the new
  trial. This is known as \emph{the meta-analytic-predictive (MAP)}
  prior~\cite{Spiegelhalter2004,Neuenschwander2010,Schmidli2014} 
\begin{equation}
\label{map:prior}
\theta_{\star} \vert Y_1,\ldots,Y_J
\end{equation}
For the NNHM with known $\tau$
\begin{equation}
\label{NNHM:thetastar}
\theta_{\star} \vert Y_1,\ldots,Y_J, \tau \sim N(\hat{\mu},\tau^2+1/w_+)
\end{equation}
which follows from (\ref{NNHM:theta}) by adding the new trial (with no
data) to the model, i.e., $s_{\star}=\infty$ and $B_{\star}=1$.

\subsection*{Analysis for new trial }
Eventually, after the new data $Y_{\star}$ have been
observed, inference for $\theta_{\star}$ can be done in two ways:
\begin{itemize}[leftmargin=11.0mm]
\item[\MAP] the \emph{meta-analytic-predictive (MAP)} approach
  formally combines the prior (\ref{map:prior}) with
  $\Ystar$ in a standard Bayesian way.
\item[\MAC] the \emph{meta-analytic-combined (MAC)} approach does not require a prior
distribution for $\thstar$. It simply infers $\thstar$ at the end of the new
trial by a meta-analysis of historical and new data, resulting in
\begin{equation}
\label{mac:prior}
\thstar \vert Y_1,\ldots,Y_J,\Ystar
\end{equation}
\end{itemize}
Importantly, \emph{MAC} and \emph{MAC} give identical
  results~\cite{Schmidli2014}.  The \emph{MAP} approach is technically more involved
  because \emph{MAP} priors (\ref{map:prior}) do not follow standard distributions and are typically heavy-tailed.
  This complicates the Bayesian analysis with $Y_{\star}$ at the end
  of the trial, which can be addressed via mixture approximations~\cite{Schmidli2014}. However, even if a \emph{MAC} analysis will
  usually be the method of choice and easy to perform with
  meta-analytic software, the \emph{MAP} prior plays
  an important role: it quantifies prior information at the design
  stage, which may be required in the trial protocol.

\subsection*{Effective sample sizes}
\label{sec:methods:ess}
In many applications, the appropriate use of prior information will
lead to smaller trials. The amount of information is ideally expressed
as an equivalent approximate prior \emph{effective sample size (ESS)}. 
In our setting we are interested in $ESS_{\star}$,
the prior effective sample size of the MAP prior
(\ref{map:prior}). Various approaches to $ESS$ have been
proposed~\cite{Neuenschwander2010, Malec2001, Pennello2008, Morita2008, Hampson2014}; they are similar in the sense that they relate
the $ESS$ to the precision (inverse of variance) of the prior
distribution.\\[2\baselineskip]
Here, we will use an approximate two-variances approach which requires: the
variance $V_{\star}$ of the analysis of interest, for which the
$ESS_{\star}$ is unknown; and, the variance $V_0$ of a simpler
analysis (e.g. a meta-analysis with $\tau=0$) with
known $ESS_0$. Assuming that effective sample sizes are approximately
proportional to precisions, the ESS of interest is
\begin{equation}
ESS_{\star} = ESS_0 \times \frac{V_0}{V_{\star}}
\end{equation}
In our case, $V_{\star}$ will be variance of the $MAP$ prior
(\ref{map:prior}), whereas $V_0$ will be the one from
the analysis assuming no between-trial heterogeneity ($\tau=0$).\\[2\baselineskip]

\section*{Case Study}
\label{sec:application}
We now illustrate a design which utilizes phase II data for the design and analysis of a phase III study. The design relies on the methodology of Section 2 and additional considerations such as practical feasibility and regulatory requirements. Data from three phase II and one phase III trial on Zirgan (0.15\% gel) for the treatment of acute herpetic keratitis will be used in the case study. All analyses were conducted in \texttt{R}~\cite{Rsoftware} with the package \texttt{bayesmeta}~\cite{bayesmeta} (see Appendix for code).\\[2\baselineskip]

\subsection*{Background}
Herpetic keratitis is an inflammatory condition of the eye caused by an outbreak of the herpes simplex virus (HSV)\cite{Kay2006,White2014}. It can have serious consequences and remains the leading cause of corneal blindness in the industrialized world \cite{Dawson1976, Suresh1999}. With as few as 1.5 million people affected world-wide~\cite{Farooq2012}, it has been classified as an orphan indication by the FDA~\cite{CorrespFDA} and the European Medical Agency~\cite{EMAOrphan2008}. \\[2\baselineskip]

In 2009, the FDA approved Zirgan for the treatment of herpetic keratitis (dendritic ulcers)~\cite{ApprovalFDA2009}. To discuss all details of the approval is beyond the scope of this application (see the publicly available documents~\cite{ApprovalDocsFDA}). However, a few points are noteworthy. Most importantly, from the files~\cite{CorrespFDA,ApprovalDocsFDA} it appears that approval was based on a retrospective analysis of the four relevant studies, three phase II and one phase III study. Retrospective means that the sponsor submitted the results of the studies after they were conducted, rather than seeking the agency's advice beforehand. Subsequently, this led to discrepancies between the sponsor's and FDA's primary analyses, including changes of the population, of the endpoint and from superiority to non-inferiority.\\[2\baselineskip]

The reasons behind this rather unusual approach to approval are not entirely clear. One explanation may be that the original manufacturer (Th\'{e}a of France) did not intend to bring Zirgan to the US market on its own; rather, it sold the license for the US market to Sirion Therapeutics in 2007 which then initiated the submission. This and the fact that the clinical studies were already conducted in the 1990s may explain why no early discussions with the FDA took place.\\[2\baselineskip]

Our goal here is not to reconstruct the approval history in detail. Rather, we will use the example to discuss an alternative, more efficient statistical approach towards approval, based on the following design specifications in the non-inferiority setting: cure rate at day 14 as endpoint, dendritic and geographic ulcers as population, and an absolute non-inferiority margin of 12 percentage points. Furthermore, we will use the risk ratio (RR) to quantify the treatment effect.\\[2\baselineskip] 

In the following, we present the evidence available at the hypothetical end-of-phase II meeting, a potential phase III trial and approval strategy, and the results of the actual phase III trial.

\subsection*{Hypothetical end-of-phase II meeting}
Three randomized phase II studies~\cite{ZirganClinPaper} were conducted between April 1990 and October 1992 (Table~\ref{tbl:data}). The studies were similar, with the only minor difference being the treatment regimen in study 6. For simplicity, we assume that this difference is not relevant for the clinical outcome.\\[2\baselineskip]

We now turn the clock back and assume we are in the situation of an end-of-phase II meeting. We assume that the sponsor would agree to a non-inferiority analysis of Zirgan versus Acyclovir (the standard of care) with the primary endpoint being cure rate at day 14. Actually, setting a non-inferiority margin proved to be difficult. For cure rate at day 14, the FDA determined two effect sizes M1~\cite{FDAGuidanceNI}: 14\% and 18\%~\cite{ApprovalDocsFDA}. The latter implies an absolute non-inferiority margin of 12 percentage points when retaining one third of the effect. We assume here that this margin had been agreed to.\\[2\baselineskip]

At this stage, it is interesting to perform a non-inferiority analysis (Zirgan versus Acyclovir) of the phase II data. If the evidence were overwhelming, it would be fair to ask whether a phase III study were required, or if approval could be granted based on the phase II data only.\\[2\baselineskip]

Our interest is the phase III treatment effect. However, since no phase III data are available yet, the phase III treatment effect corresponds to the predicted treatment effect $\theta_{\star}$ from the phase II studies (see Section 2). The underlying statistical model is the NNHM~\eqref{EX},~\eqref{EX:normal}, with study-specific estimates of the log-risk-ratios $Y_j = \log(\text{RR}_j)$ and standard errors 

\begin{equation}
s_j = \sqrt{1/r_C - 1/n_C + 1/r_T - 1/n_T}
\end{equation}

 where $n$ and $r$ denote the number of patients and responders. This requires a transformation to the risk difference scale and a sensible prior distribution for the between-trial heterogeneity parameter $\tau$ (the prior for $\mu$ will be non-informative).\\[2\baselineskip]

The first point is straightforward. For a response rate $p_C$ in the control group and a pre-defined non-inferiority margin $m=p_C - p_T$, the transformation is given by the definition of the risk ratio; for $p_C = 0.9$ (the assumed cure rate for Acyclovir based on historical data) and margin $m = 0.12$, non-inferiority holds if $\text{RR}_{T:C} \geq 0.867$.\\[2\baselineskip]

For the $\tau$ prior we use $\tau \sim HN(0.5)$, which has median 0.34 and 95\% interval (0.016;1.12). This prior is centered at moderate to substantial heterogeneity and covers small to large heterogeneity (see Table~\ref{tab:heterogeneity}). Notably, one may perform a sensitivity analysis using a prior which favors larger between-trial heterogeneity, e.g. $\tau \sim HN(1)$.\\[2\baselineskip]

\begin{figure}[h!]
    \caption{Data and results at end-of-phase II meeting}
    \centering
    \includegraphics[width=\textwidth]{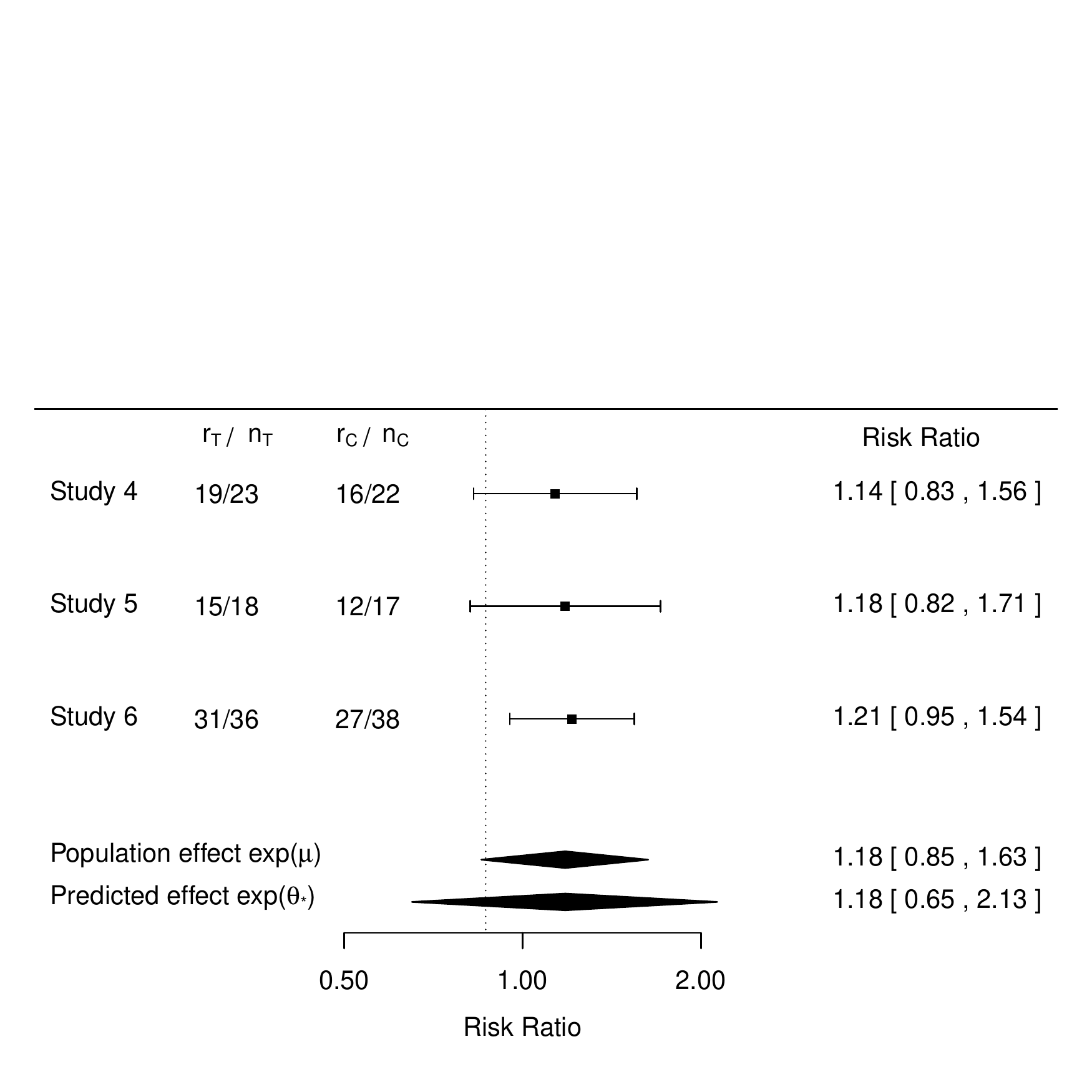}
    \label{fig:eop2_ma}
\end{figure}

The meta-analysis of the phase II data is shown in Figure~\ref{fig:eop2_ma}, where the data, study-specific (stratified) risk ratios $\text{RR}_j$, the  population mean $\mu$ and the predicted effect $\theta_{\star}$ are shown. The reference line is drawn at the non-inferiority margin ($\text{RR}_{T:C}=0.867$ for $p_C = 0.9$). The posterior for $\tau$ indicates small between-trial heterogeneity, with median 0.12 (95\% interval 0.00 to 0.51). \\[2\baselineskip]

The meta-analysis provides evidence for non-inferiority. If $\mu$ were the parameter of interest, an almost conclusive statement would follow: the lower bound of the 95\% interval is just below the non-inferiority margin. In fact, $P(\mu \geq \log(0.867)) = 97.1\%$, very close to $97.5\%$. However, the parameter $\theta_{\star}$ in the phase III trial is of interest. For this parameter, the evidence for non-inferiority is weaker, but still substantial: $P(\theta_{\star} \geq \log(0.867)) = 92.0\%$. 

\subsection*{Phase III study and proposed strategy for approval}
Designing a phase III study that allows to assess non-inferiority in combination with the available evidence is desirable. Not only will this allow to run a smaller study, it will also provide a treatment effect estimate based on all relevant evidence. However, regulators may have good reasons to argue that a smaller study may provide insufficient information for approval, especially to assess the safety and risk/benefit ratio.\\[2\baselineskip]

We now discuss the design of  a phase III study (study 7) which uses phase II data and allows for seeking approval based on an interim analysis. Depending on negotiations with regulators, a post-approval commitment to run the study to its end (even if approval is granted at interim) may be required. However, such negotiations will always be case-specific, highlighting the importance of early discussions with regulators. Nevertheless, the option to seek approval based on a positive interim analysis seems attractive for this case study. Since the endpoint is evaluated at day 14, there will be a small time window between the last patient enrolled for the interim analysis and the actual data read-out and analysis. With an anticipated recruitment period of two years, such a strategy could result in a markedly earlier approval. \\[2\baselineskip]

When seeking approval based on interim results, the information fraction for the interim analysis becomes a key design aspect. We will assume that the interim analysis is conducted after 50\% of the patients have been evaluated. For the sample size, in order to align with the actual study as originally conducted, we will assume $n_C = n_T = 80$. This results in interim sample sizes $\tilde{n}_{C} = \tilde{n}_{T} = 40$.\\[2\baselineskip]

It is also important to understand how much phase II information is borrowed (which depends on the between-trial heterogeneity) when inferring the phase III effect. Using the variance ratio approach (Section 2), the ESS is 14. 

\subsection*{Operating characteristics}
We evaluate the operating characteristics (type I error rate and power) of the design and compare them to a phase III design ignoring the phase II data. The operating characteristics are presented in Table~\ref{tbl:oc} based on 10'000 simulations conducted in \texttt{R}~\cite{Rsoftware} with the package \texttt{bayesmeta}~\cite{bayesmeta}. For different response rates $p_C$ and treatment differences $\delta$, two probabilities are shown: the probability to be successful at the final analysis (regardless of the outcome of the interim analysis), and the probability to be successful both at the interim and the final analysis.\\[2\baselineskip]

The gain in power for the proposed design can be substantial. For example, for $p_C = 0.7$ (the Acyclorivr cure rate observed in phase II) and $\delta = 0.06$, the power is 87\% versus 66\%. The power gain is even larger at interim (70\% versus 35\%). When $p_C = 0.9$ (the observed cure rate for Acyclovir based on historical data) and $\delta = 0$, the power is 87\% versus 79\%, and 68\% versus 48\% at interim. The larger increase in power at interim is remarkable and due to the highly consistent phase II results, which suggested superiority of Zirgan.\\[2\baselineskip]

The gain in power, however, comes at the price of an increased type I error rate. Strict type I error rate control cannot be guaranteed~\cite{Viele2014}. For example, for $p_C = 0.7$ and $p_C = 0.9$, the type I error rates are 6\% versus 1\% and 8\% versus 3\%. This increase is not dramatic, yet it cannot be ignored and needs to be discussed with regulators during the design phase. If it is of concern, robust prior distributions could be considered~\cite{Schmidli2014}.

\subsection*{Actual phase III data and analysis}
\begin{figure}
    \caption{Data and results for interim and final analysis in Phase III}
    \centering
    \includegraphics[width=\textwidth]{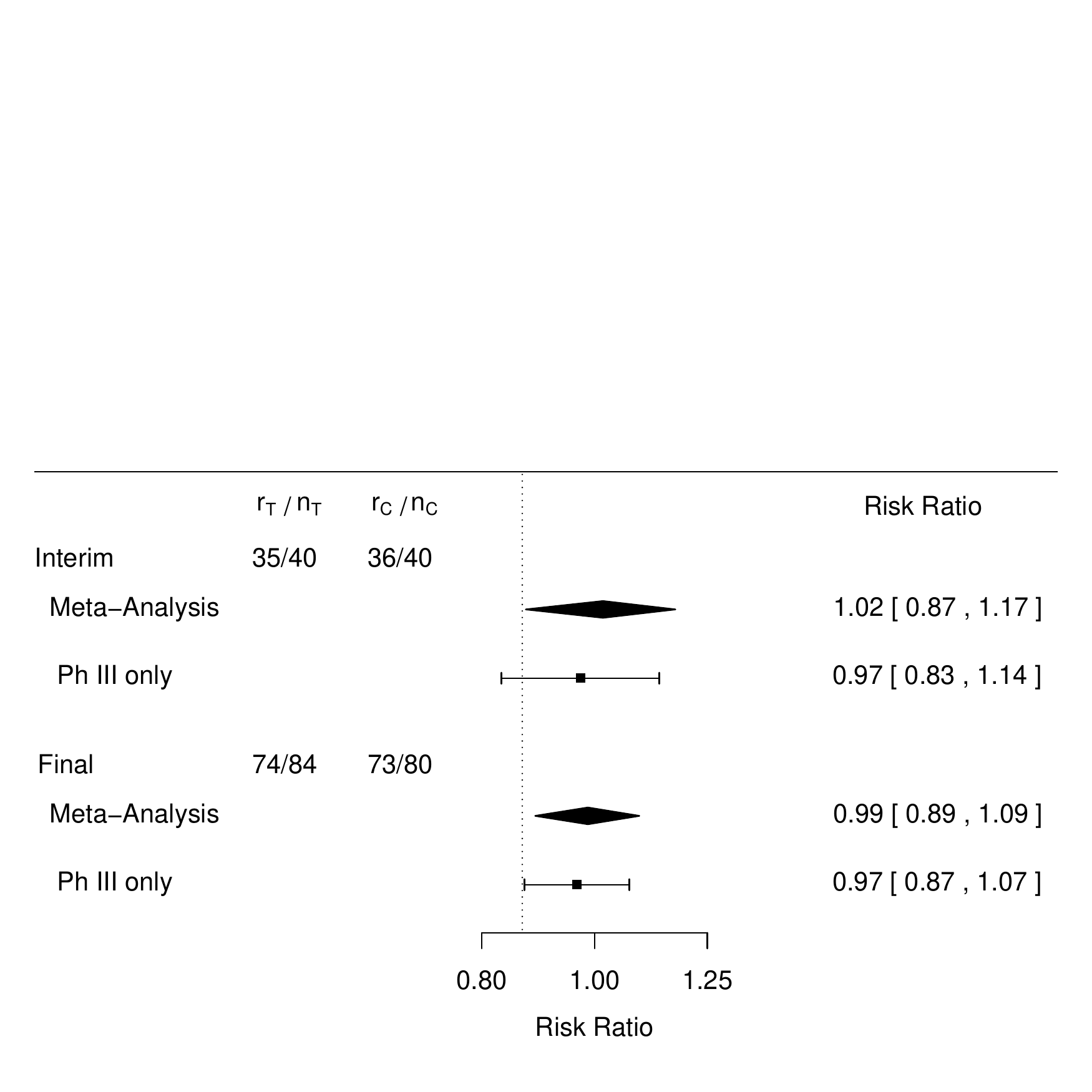}
    \label{fig:final_analysis}
\end{figure}

The actual data observed in the phase III study are only available for the final analysis. In order to reconstruct an interim analysis using half of the patients, we use an interim sample size of 40 per arm. Furthermore, we choose the number of responders such that observed response rate at interim is close to the observed response rate at the final analysis (see Figure~\ref{fig:final_analysis}).\\[2\baselineskip]

The results are presented in Figure~\ref{fig:final_analysis}. The interim analysis based on all data (meta-analysis) allows to declare non-inferiority. Note that non-inferiority is claimed based on the parameter corresponding to study 7 ($\theta_{\star}$) incorporating the evidence from studies 4, 5 and 6. On the other hand, the evidence from the phase III study alone is insufficient to declare non-inferiority at interim because the 95\% interval includes the non-inferiority margin.\\[2\baselineskip]

As mentioned before, the idea would be to gain approval with the interim phase III data supported by phase II via the meta-analysis, assuming other data (such as safety) is also favorable. Yet, depending on negotiations with regulators, the study may still run to its end, allowing a more robust evaluation of the effect at the final analysis. The results for the final analysis are also shown in Figure~\ref{fig:final_analysis}. For the meta-analysis, the interval for the risk ratio becomes narrower and still excludes the non-inferiority margin, thus confirming the interim result. The analysis using the phase III study leads to a lower bound of the interval ($0.870$) which is just above the non-inferiority threshold $0.867$, also allowing to conclude non-inferiority.\\[2\baselineskip]

Finally, results for $\tau$ indicate small between-trial heterogeneity at the interim and the final analysis. The posterior median (95\% interval) is 0.12 (0.00 to 0.41) for the interim and  0.13 (0.00 to 0.43) for the final analysis. This supports the consistency of the results across all studies.\\[2\baselineskip]

\section*{Discussion}
\label{sec:discussion}
Here we presented a simple, yet attractive design in rare diseases using phase II data in phase III studies. We illustrated it for binary endpoints, but the extension to other endpoints is straightforward.\\[2\baselineskip]
The proposed approach uses the phase II data prospectively, which has obvious advantages. First, fewer patients are required in the phase III study. Second, the estimate combines all available evidence. And third, due to the nature of the approach, extreme results will be pulled towards the population mean. The Zirgan case study used to illustrate the design is built on real data as submitted to the FDA. However, the FDA approved Zirgan for a different indication (dendritic ulcers only) and endpoint (cure rate at day 7) than those used in our case study. \\[2\baselineskip]

Of course, as with any design, all stakeholders need to be convinced. It may be argued that the case study is quite atypical since phase II studies are often not randomized in orphan diseases. This, however, becomes a self-fulfilling prophecy: if evidence from randomized phase II studies is only considered supportive, there is little motivation to perform them. On the other hand, if data from randomized phase II studies could be used, this would make them more attractive. It is therefore important that patient groups, regulators and sponsors consider such designs.\\[2\baselineskip]

Other designs have been proposed before, and an excellent overview is given in Korn et al~\cite{Korn2013}. Some have been implemented in practice, for example the historical control monotherapy design proposed by French et al~\cite{French2010}. This design was used successfully, resulting in the approval of Aptiom (eslicarbazepine acetate) for the treatment of partial-onset seizures~\cite{ApprovalFDA2015Aptiom,LabelFDA2015Aptiom}. Other examples include N-of-1 trials~\cite{Guyatt1986}, global studies~\cite{Kuerner2015}, or basket trials, e.g. the B2225 study for Imatinib~\cite{Heinrich2008}.\\[2\baselineskip]

It is also worth mentioning that recent initiatives to improve the drug development process send encouraging signals that a better use of the evidence is welcomed. Important directions are given in the 21\textsuperscript{st} Century Cures Act~\cite{21CuresAct}, which encourages the FDA to further evaluate the use of Bayesian methodology and non-randomized evidence. Furthermore, calls have been made to make the drug approval process more continuous and flexible to account for evidence as it accumulates~\cite{Roy2012}. The European Medicines Agency has also initiated various working groups.\\[2\baselineskip]

It is clear that we only considered a small portion of the drug approval process. Efficacy plays a unique role when seeking approval, but other measures are also important. Safety is critical, and additional evidence may be required to assess long-term risks. However, this can often be achieved as a post-approval requirement in the form of non-randomized open-label studies. This approach has the advantage that patients have early access to the treatment whilst additional data are collected. \\[2\baselineskip]

The proposed approach has limitations. The potential increase in type I error needs to be considered and may require design modifications, including robust meta-analytic models~\cite{Schmidli2014}. Likewise, for a non-inferiority design, one may consider to directly model the risk difference and use a meta-analytic approach on this scale (e.g. Warn et al~\cite{Warn2002}). However, most applications will be superiority trials, for which relative measures such as risk ratios or odds ratios are common. Finally, we did not use historical data to inform the prior for the between-trial heterogeneity ($\tau$), even though this would be possible~\cite{Turner2015}.\\[2\baselineskip]

The motivation of this paper was not to challenge FDA's decision. On the contrary: only due to the many publicly available FDA documents, we were able to use this insightful example. We hope that it will facilitate the implementation of the proposed design in practice.

\clearpage
\section*{Declaration of Conflicting Interest}
Dr. Wandel and Dr. Neuenschwander are employed by Novartis Pharma AG, Basel, Switzerland.

\clearpage
\textbf{Tables}

\begin{table}[h!]
\centering
\caption{
  Classification of between-trial heterogeneity with 97.5\% quantile to median
  ratio for risk ratio (RR) trial parameters;  $\sigma$ is the
  outcome standard deviation, $\tau$ is the between-trial standard
  deviation.}
\vspace*{\baselineskip}
\begin{tabular}{ccc}
heterogeneity ($\sigma/\tau$) & $\tau$ (if $\sigma=2$) &  
$\text{RR}_{97.5\%}/\text{RR}_{50\%}$ 
\\ \hline
large (2)& 1 &  7.10 \\ 
substantial (4) & 0.5 & 2.66  \\
moderate (8) & 0.25 &  1.63 \\
small (16) & 0.125 &  1.28 \\
\end{tabular}
\label{tab:heterogeneity}
\end{table}

\begin{sidewaystable}
\caption{Data of Phase II and III studies}
\vspace*{\baselineskip}
\label{tbl:data}
\begin{tabular}{l|cccc}
\hline
   & \multicolumn{4}{c}{Study (Phase)}\\
   & 4 (II) & 5 (II) & 6 (II) & 7 (III)\\
\hline
Objective & Efficacy \& Safety & Efficacy \& Safety & Efficacy \& Safety & Efficacy \& Safety\\
Design & 3-arm randomized & 2-arm randomized & 3-arm randomized & 2-arm randomized\\
Location & Africa & Europe & Pakistan & Europe \& Africa\\
Product  & G: 0.15\%, 0.05\%; A: 3\% &G: 0.15\%; A: 3\% &G: 0.15\%, 0.05\%; A: 3\% &G: 0.15\%; A: 3\% \\
Regimen & 1 & 1  & 2 & 1\\
Study period (months) & 4/90--5/92 (25) & 12/90--5/92 (18) & 5/91--10/92 (18) & 9/92--9/94 (25)\\
Total cure rate, day 14 (\%) &  & & &\\
\qquad Zirgan & 19/23 (82.6) & 15/18 (83.3) & 31/36 (86.1) & 74/84 (88.1)\\
\qquad Acyclovir & 16/22 (72.7) & 12/17 (70.6) & 27/38 (71.1) & 73/80 (91.3)\\
\hline
\multicolumn{5}{p{250pt}}{}\\
\multicolumn{5}{p{540pt}}{Regimen: 1 = 1 drop 5x/day until ulcer healed, then 1 drop 3x/day for 7 days; 2 = 1 drop 5x/day for 10 days}
\end{tabular}
\end{sidewaystable}

\begin{table}
\caption{Operating characteristics for phase II/III (meta-analysis) and phase III alone}
\vspace*{\baselineskip}
\label{tbl:oc}
\setlength{\tabcolsep}{4pt}
\begin{tabular}{c|ccccc}
             & \multicolumn{5}{c}{$\delta = p_T - p_C$}\\
 $p_C$  & -0.12  & -0.06 &  0.0 & 0.06 & 0.12\\
\hline
  0.70 &  6 (3) &  25 (15)             &  56 (39) &  $\;$ 87 (70)  &  $\;$ 98 $\;$ (90)\\ 
          &  1 (0) &  $\;$ 8 $\;$ (3)  &  30 (12) &  $\;$ 66 (35)  &  $\;$ 93 $\;$ (67)\\ 
\rule{0pt}{3ex}
  0.75 &  7 (4) &  26 (16)             &  61 (44) &  $\;$ 91 (75)  &  100 $\;$ (94)\\ 
          &  1 (0) &  10 $\;$ (4)        &  36 (16) &  $\;$ 76 (44)  &  $\;$ 98 $\;$ (78)\\ 
\rule{0pt}{3ex}
  0.80 &  7 (4) &  29 (18)             &  68 (49) &  $\;$ 94 (80)  &  100 $\;$ (97)\\ 
          &  2 (1) &  13 $\;$ (5)       &  46 (22)  &  $\;$ 87 (57)  &  100 $\;$ (90)\\ 
\rule{0pt}{3ex}
  0.85 &  7 (4) &  32 (19)             &  76 (55) &  $\;$ 98 (88)  &  100 (100)\\ 
          &  3 (1) &  17 $\;$ (7)       &  60 (31)  &  $\;$ 95 (72)  &  100 $\;$ (99)\\ 
\rule{0pt}{3ex}
  0.90 &  8 (4) &  38 (24)            &  87 (68)  &  100 (98)        &  --\\ 
          &  3 (1) &  26 (11)             &  79 (48) &  100 (94)        &  --\\ 
\hline
\multicolumn{6}{p{250pt}}{}\\
\multicolumn{6}{p{250pt}}{Percentages presented: probability for success at final (probability for success at interim and final). The first row corresponds to the meta-analysis, the second row to the analysis of the phase III study alone.}
\end{tabular}
\end{table}

\clearpage
\textbf{Figures}

\newpage

\clearpage
\begin{appendix}
\section*{Appendix}
\begin{verbatim}
library("bayesmeta")
library("metafor")

################################################ 
#  Part I - ratio of RRs for tau ~ HN(...)
################################################ 
set.seed(314)
N <- 1000000
tau1 <- abs(rnorm(N, 0, 0.5))
tau2 <- abs(rnorm(N, 0, 1.0))
theta1 <- rnorm(N, 0, tau1)
theta2 <- rnorm(N, 0, tau2)
exp(quantile(theta1, prob = 0.975))
exp(quantile(theta2, prob = 0.975))


################################################ 
#  Part II - application 
################################################ 

# --------------------------------------------
# read-in data
# transform into 2x2 cell entries
# derive mean and se from normal approximation
# --------------------------------------------

all.data <- data.frame(study = c("4", "5", "6", "7IA", "7FA"),
                       rt    = c(19, 15, 31, 35, 74),
                       nt    = c(23, 18, 36, 40, 84),
                       rc    = c(16, 12, 27, 36, 73),
                       nc    = c(22, 17, 38, 40, 80))

all.data$ai <- all.data$rt
all.data$bi <- all.data$nt - all.data$rt
all.data$ci <- all.data$rc
all.data$di <- all.data$nc - all.data$rc

nmappr <- escalc(ai=ai, bi=bi, ci=ci, di=di,
                 data=all.data,	measure="RR")

# --------------------------------------------------
# analyses:
#  - end-of-phase-II
#  - interim for phase III
#  - final for phase III
# summaries: 
#  - population mean,
#  - predicted effect for end-of-phase-II
#  - phase-III effect for interim and final analyses
# --------------------------------------------------

eop2  <- bayesmeta(y = nmappr$yi[1:3], sigma = sqrt(nmappr$vi[1:3]),
                   tau.prior = function(t){dhalfnormal(t,scale=0.5)})

ph3IA <- bayesmeta(y = nmappr$yi[1:4], sigma = sqrt(nmappr$vi[1:4]),
                   tau.prior = function(t){dhalfnormal(t,scale=0.5)})

ph3FA <- bayesmeta(y = nmappr$yi[c(1:3,5)], sigma = sqrt(nmappr$vi[c(1:3,5)]),
                   tau.prior = function(t){dhalfnormal(t,scale=0.5)})

round(exp(t(eop2$summary)[,c("median", "95% lower", "95% upper")]), 3)
round(exp(ph3IA$theta[c("median", "95% lower", "95% upper"),4]), 3)
round(exp(ph3FA$theta[c("median", "95% lower", "95% upper"),4]), 3)
\end{verbatim}
\end{appendix}

\end{document}